\documentclass[twocolumn,showpacs,preprintnumbers,amsmath,amssymb,pre]{revtex4}

\usepackage{graphicx}
\usepackage{dcolumn}
\usepackage{bm}
\usepackage{enumerate}
\usepackage{amsmath}
\usepackage{color}

\begin{document}

\title{First passage time for $g$--subdiffusion process of vanishing particles}

\author{Tadeusz Koszto{\l}owicz}
 \email{tadeusz.kosztolowicz@ujk.edu.pl}
 \affiliation{Institute of Physics, Jan Kochanowski University,\\
         Uniwersytecka 7, 25-406 Kielce, Poland}

\date{\today}

\begin{abstract}
Subdiffusion equation and molecule survival equation, both with Caputo fractional time derivatives with respect to another functions $g_1$ and $g_2$, respectively, are used to describe diffusion of a molecule that can disappear at any time with a constant probability. The process can be interpreted as ``ordinary'' subdiffusion and ``ordinary'' molecule survival process in which timescales are changed by the functions $g_1$ and $g_2$. We derive the first-passage time distribution for the process. The mutual influence of subdiffusion and molecule vanishing processes can be included in the model when the functions $g_1$ and $g_2$ are related to each other. As an example, we consider the processes in which subdiffusion and molecule survival are highly related, which corresponds to the case of $g_1\equiv g_2$.
\end{abstract}

\maketitle

{\it Introduction.} Subdiffusion occurs in media in which the movement of molecules is very hindered due to a complex structure of a medium. Within the Continuous Time Random Walk model the distribution of waiting time for a molecule to jump has a heavy tail, $\tilde{\psi}(\tau)\sim 1/\tau^{\alpha+1}$, where $\alpha\in(0,1)$ is a subdiffusion parameter \cite{mk,mk1,ks}. The parameter is constant in a homogeneous system in which diffusion properties do not change with time. The process can be described by the subdiffusion equation with the ``ordinary'' Caputo fractional time derivative. Changing time scale may cause subdiffusion to be accelerated or delayed. An example is the subordinate method, in which a stochastic process changing this scale is involved \cite{sokolov1,csm,dybiec,stan2020}. Recently, it has been proposed the subdiffusion equation with a fractional time Caputo derivative with respect to another function $g$ ($g$--subdiffusion equation) \cite{kd2022,kd2021a,kd2021b}. The deterministic function $g$ rescales the time variable in the ``ordinary'' subdiffusion model. For example, the waiting time for the molecule to jump in $g$--subdiffusion model is described by the $\psi$ distribution which is related to $\tilde{\psi}$ as $\psi(\tau)=\tilde{\psi}(g(\tau))$ \cite{kd2021b}. 

We study subdiffusion of a molecule that can be eliminated at any moment from further diffusion, the molecule can be retained permanently or decayed; we treat both processes as "vanishing" of the molecule. We assume that the probability density of a molecule disappearing does not depend on a molecule position. Molecule elimination and diffusion processes can be related to each other. An example is diffusion of an antibiotic through a bacterial biofilm. Bacteria have different defense mechanisms against the action of the antibiotic, see \cite{aot,mot} and the references cited in \cite{km,kmwa}. One of them is biofilm compaction, which makes it difficult for antibiotic molecules to diffuse and even stops them. The permanent arrest of the molecule can be treated as its "disappearance" because it is eliminated from further diffusion. Since the antibiotic mainly attacks rapidly growing bacteria, another defense mechanism is to temporarily stop the bacteria from multiplying. This mechanism even facilitates diffusion of the antibiotic through the biofilm. Another example is subdiffusion of antibiotic molecules in a medium having a "plum pudding" structure. This process can occur when a person infected with cystic fibrosis is also infected with {\it Pseudomonas aeruginosa} bacteria \cite{kmwa}. The "pudding" represents the cystic fibrosis mucus, and the "plums" represent the {\it Pseudomonas aeruginosa} biofilm. The purpose of the antibiotic is to kill the {\it Pseudomonas aeruginosa} bacteria whose defense mechanisms can destroy the antibiotic molecules. In this case, slowing subdiffusion in the "pudding" may increase the chance of avoiding the decay of the antibiotic molecules. These examples show that the processes of diffusion and decay of molecules can be related to each other. Slowing down molecules diffusion can cause an increase as well as a decrease of their disappearance probability. Thus, relations between diffusion and processes that lead to a molecules disappearance may be unique to a given process. Later we will pay attention to the process in which subdiffusion and molecule survival are highly related, and transport properties of a medium change periodically.   

We consider the following problem. A molecule is located at point $x_0$ at the initial time $t=0$ in a one--dimensional homogeneous system. We find the probability distribution of time that a molecule reaches the point $x_M$ for the first time. This problem is potentially of great importance in modeling of the transport of antibiotics in bacterial biofilms and of epidemic spreading when the arrival of an infected individual to some point may create a new source for the spread of the epidemic. 

{\it Model.} We assume that the disappearance of a molecule is independent of its position. Let $P(x,t|x_0)$ be a probability density (Green's function) that the molecule is at the point $x$ at time $t$ when particle decay would be ``turned off'', $x_0$ is the initial molecule position, and let $\rho(t)$ be a probability that the molecule still exists at time $t$. Then, the Green's function $P_\rho$ for the vanishing molecule reads
\begin{equation}\label{eq1}  
P_\rho(x,t|x_0)=\rho(t)P(x,t|x_0).
\end{equation}

We also assume that in some initial time interval a diffusion process is ``ordinary'' subdiffusion and the molecule vanishing process is described by the ``ordinary'' decay equation. The ``ordinary'' subdiffusion is described by the following equation 
\begin{equation}\label{eq2}
\frac{^C \partial^{\alpha} P(x,t|x_0)}{\partial t^\alpha}=D\frac{\partial^2 P(x,t|x_0)}{\partial x^2},
\end{equation}
where $^C\partial^\alpha f(t)/\partial t^\alpha=\int_0^t (t-u)^{-\alpha}f'(u)du/\Gamma(1-\alpha)$ is the ``ordinary'' Caputo fractional derivative, $0<\alpha<1$, $f'(t)=df(t)/dt$, $D$ is a subdiffusion coefficient given in the units of $m^2/s^\alpha$. 

A frequently used equation to describe the disappearance of a molecule is $d\phi(t)/dt=-\lambda\phi(t)$, where $\phi$ is a probability density that molecule decays at time $t$ and $\lambda$ is a decay rate. The solution to this equation reads $\phi(t)=\lambda{\rm e}^{-\lambda t}$. The probability that the molecule still exists at time $t$ is $\rho(t)=1-\int_0^t \phi(u)du={\rm e}^{-\lambda t}$. This function fulfils the molecule survival equation (MSE) $d\rho(t)/dt=-\lambda\rho(t)$ with the initial condition $\rho(0)=1$. However, if the function $\rho$ has a heavy tail, $\rho(t)\sim 1/t^\beta$ with $\beta\in(0,1)$ when $t\rightarrow\infty$, the MSE involves ``ordinary'' fractional Caputo derivative of the order $\beta$ \cite{suppl}
\begin{equation}\label{eq3}
\frac{^C d^\beta \rho(t)}{dt^\beta}=-\lambda\rho(t),
\end{equation}
the unit of $\lambda$ is $1/s^\beta$. 

We assume that at some initial stage the process is described by Eqs. (\ref{eq2}) and (\ref{eq3}). In the further course of the process subdiffusion and the molecule vanishing process may change due to the influence of external factors and/or due to the interaction of the processes. In order to describe the processes in the whole time domain, we assume that they are described by equations in which ``ordinary'' Caputo derivatives are replaced by $g$--Caputo derivatives, see Eq. (\ref{eq4}) presented below. However, for short time the new equations should take the form of Eqs. (\ref{eq2}) and (\ref{eq3}). 

The $g$-Caputo fractional derivative $^Cd^{\alpha}_g f(t)/dt^\alpha$ of the order $\alpha$ with respect another function $g$ is defined for $0<\alpha<1$ as \cite{almeida} 
\begin{equation}\label{eq4}
\frac{^Cd^{\alpha}_g f(t)}{dt^\alpha}=\frac{1}{\Gamma(1-\alpha)}\int_0^t (g(t)-g(u))^{-\alpha}f'(u)du,
\end{equation}
where $f'(t)=df(t)/dt$, the function $g$ fulfils the conditions $g(0)=0$, $g(\infty)=\infty$, and $g'(t)>0$ for $t>0$. The values of function $g$ are given in a time unit. When $g(t)=t$, the $g$-Caputo fractional derivative takes the form of the ``ordinary'' Caputo derivative.
For $\alpha=1$, the $g$--Caputo derivative takes the form of the Riemann--Stieltjes derivative $^Cd^{1}_g f(t)/dt^\alpha\equiv f^{[1]}_g=f'(t)/g'(t)$. 

We involve the $g$--Caputo fractional time derivative in both the subdiffusion equation and the molecular survival equation. We assume that $g$--subdiffusion is controlled by a function $g_1$, and the $g$--molecule survival process by a function $g_2$, 
\begin{equation}\label{eq5}
\frac{^C \partial^{\alpha}_{g_1} P(x,t|x_0)}{\partial t^\alpha}=D\frac{\partial^2 P(x,t|x_0)}{\partial x^2},
\end{equation}
\begin{equation}\label{eq6}
\frac{^C d^\beta_{g_2} \rho(t)}{dt^\beta}=-\lambda\rho(t),
\end{equation}
$0<\alpha,\beta\leq 1$.
As we have assumed, Eqs. (\ref{eq5}) and (\ref{eq6}) describe the process over the entire time domain, but in the short time these equations take the form of Eqs. (\ref{eq2}) and (\ref{eq3}), respectively. The latter condition is fulfilled if $g_i(t)\rightarrow t$ when $t\rightarrow 0$. We assume that $g_i(t)=t+\xi_i(t)$ with $\xi_i$ satisfying $\xi_i(t)\rightarrow 0$ when $t\rightarrow 0$, $i=1,2$. The functions $g_1$ and $g_2$ satisfy the general conditions $g_1(0)=g_2(0)=0$, $g_1(\infty)=g_2(\infty)=\infty$, and $g_1(t),g_2(t),g'_1(t),g'_2(t)>0$ for $t>0$.

{\it Solutions to Eqs. (\ref{eq5}) and (\ref{eq6}).} The Green's function $P$ is a solution to Eq. (\ref{eq5}) for the initial condition $P(x,0|x_0)=\delta(x-x_0)$, where $\delta$ is the delta--Dirac function, and for the boundary conditions $P(\pm\infty,t|x_0)=0$, 
\begin{equation}\label{eq7}
P(x,t|x_0)=\frac{1}{2\sqrt{D}}f_{-1+\alpha/2,\alpha/2}\left(g_1(t);\frac{|x-x_0|}{\sqrt{D}}\right),
\end{equation}
where 
\begin{eqnarray}\label{eq8}
f_{\nu,\mu}(z;a)
=\frac{1}{z^{1+\nu}}\sum_{k=0}^\infty \frac{1}{k!\Gamma(-\nu-\mu k)}\left(-\frac{a}{z^\mu}\right)^k,
\end{eqnarray}
$a,\mu>0$. The solution to Eq. (\ref{eq6}) for the initial condition $\rho(0)=1$ is
\begin{equation}\label{eq10}
\rho(t)={\rm E}_{\beta}\big(-\lambda g^\beta_2(t)\big),
\end{equation}
where ${\rm E}_\beta(z)=\sum_{n=0}^\infty z^n/\Gamma(\beta n+1)$ is the Mittag--Leffler function, $0<\beta\leq 1$. We note that for $\beta=1$ we have $\rho(t)={\rm e}^{-\lambda g_2(t)}$. From Eqs. (\ref{eq1}), (\ref{eq7}), and (\ref{eq10}) we get
\begin{eqnarray}\label{eq11}
P_\rho(x,t|x_0)=\frac{{\rm E}_{\beta}\big(-\lambda g^\beta_2(t)\big)}{2\sqrt{D}}\\
\times f_{-1+\alpha/2,\alpha/2}\left(g_1(t);\frac{|x-x_0|}{\sqrt{D}}\right).\nonumber
\end{eqnarray}
The calculations provided the above solutions are shown in \cite{suppl}. 

{\it First passage time.}
The process of molecule vanishing is independent of its position. We assume that the correlation of the functions $g_1$ and $g_2$, if any, is not local and does not depend on the variable $x$. Let the system consist of two parts $M_1$ and $M_2$ separated by a point $x_M$, $M_1=(-\infty,x_M)$ and $M_2=(x_M,\infty)$. At the initial moment $t=0$ the molecule is at the point $x_0$ located in the region $M_1$.

The probability that the molecule leaves the region $M_1=(-\infty,x_M)$ first time in the time interval $(t,t+\Delta t)$, where $\Delta t$ is assumed to be small, is $F(t;x_0,x_M)\Delta t=\rho(t)[R(t;x_0,x_M)-R(t+\Delta t;x_0,x_M)]$. $R$ is a probability that the molecule did not leave the $M_1$ region by time $t$, $R(t;x_0,x_M)=\int_{-\infty}^{x_M} P_{abs}(x,t|x_0)dx$, where $P_{abs}(x,t|x_0)$ is the probability of finding the molecule in the region $M_1$ in a system with a fully absorbing wall located at $x_M$ \cite{redner}.
The commonly used boundary condition at the absorbing wall is $P_{abs}(x_M,t|x_0)=0$. 
The Green's function for a system with a fully absorbing wall can be found by means of
the method of images, which for $x,x_0<x_M$ gives $P_{abs}(x,t|x_0)=P(x,t|x_0)-P(x,t|2x_M-x_0)$ \cite{feller,chandra}. After calculations we get
\begin{eqnarray}\label{eq12}
R(t;x_0,x_M)=1-f_{-1,\alpha/2}\left(g_1(t);\frac{x_M-x_0}{\sqrt{D}}\right).
\end{eqnarray}
Taking the limit of $\Delta t\rightarrow 0$ we obtain
\begin{equation}\label{eq13}
F(t;x_0,x_M)=-\rho(t)\frac{dR(t;x_0,x_M)}{dt}.
\end{equation}
Eqs. (\ref{eq8}), (\ref{eq12}), and (\ref{eq13}) provide
\begin{eqnarray}\label{eq14}
F(t;x_0,x_M)=g_1'(t){\rm E}_{\beta}\big(-\lambda g^\beta_2(t)\big)\\
\times f_{0,\alpha/2}\left(g_1(t);\frac{x_M-x_0}{\sqrt{D}}\right).\nonumber
\end{eqnarray}
For $t\leq 0$ we put $F(t;x_0,x_M)\equiv 0$. 

The probability $Q(t;x_0,x_M)$ that the molecule has passed the point $x_M$ until time $t$ is calculated by means of the formula 
\begin{equation}\label{eq15}
Q(t;x_0,x_M)=\int_0^t F(u;x_0,x_M) du.
\end{equation} 
The function $Q$ will be calculated numerically. Another function describing spreading of vanishing molecules is the mean square displacement of a particle $\sigma_\rho^2(t)=\int_{-\infty}^\infty (x-\bar{x})^2 P(x,t|x_0)dx$, where $\bar{x}$ is a mean value of $x$. For the Green's function Eq. (\ref{eq11}) we get
\begin{equation}\label{eq15a}
\sigma_\rho^2(t)=\rho(g_2(t))\frac{2D}{\Gamma(1+\alpha)}g_1^{\alpha}(t).
\end{equation}
When $g_1(t)\equiv t$ and $\rho(t)\equiv 1$, , the function $\sigma_{\rho=1}^2(t)\sim t^\alpha$ is frequently used to define a kind of diffusion; subdiffusion is when $0<\alpha<1$, normal diffusion is for $\alpha=1$, and superdiffusion for $\alpha>1$.

{\it Examples.}
As examples, we consider the case of fully correlated subdiffusion and molecule vanishing processes. We assume that $g_1(t)\equiv g_2(t)\equiv g(t)$ and that the function $\xi_i$ depends on a periodic function. As $g$ we consider the following functions: $g_0(t)=t$, $g_A(t)=t(1+a{\rm sin}(\tilde{\omega}(t)t+\phi))$, $g_B(t)=t+a((\omega t)(\omega t-{\rm sin}(2\omega t))+(1-{\rm cos}(2\omega t))/2$, and $g_C(t)=t+a({\rm sin}(\omega t+\phi)-{\rm sin}(\phi))$, $a$ and $\omega$ are positive parameters. Since $g(t),g'(t)>0$ we assume $a<1$ in the cases of $A$ and $B$, and $a<1$, $a\omega<1$ in the case of $C$. In the case of $A$ the function $\tilde{\omega}$ depends on $t$, for the numerical calculations we put $\tilde{\omega}(t)=t/(\tau_0+t)$, where $\tau_0$ is a parameter given in a time unit.

\begin{figure}[htb]
\centering{%
\includegraphics[scale=0.4]{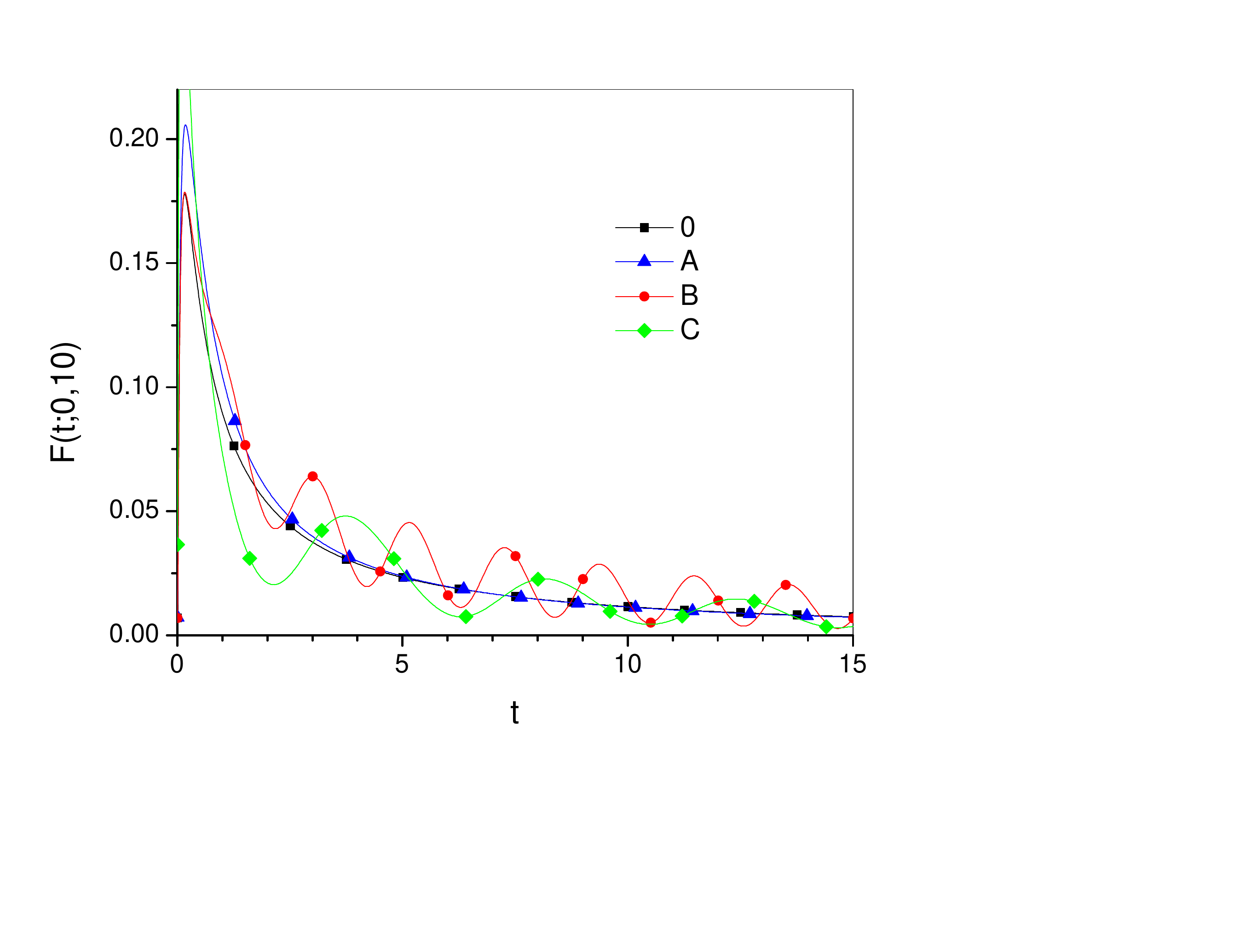}}
\caption{Time evolution of the first-passage time distribution for $g$--subdiffusion with molecule vanishing processes controlled by the functions $g_1(t)\equiv g_2(t)\equiv g(t)$, where $g=g_0$ (symbol $O$ in the legend), $g=g_A$ (symbol $A$), $g=g_B$ (symbol $B$), and $g=g_C$ (symbol $C$). The parameters are $\alpha=0.6$, $\beta=0.5$, $\lambda=0.01$, $\omega=1.5$, $\phi=0$, $a=0.4$, $x_0=0$, and $x_M=10$, $\tilde{\omega}=1.5t/(1+t)$ for $g_B$, $\tau_0=1$, all quantities are given in arbitrarily chosen units.} 
\label{fig1}
\end{figure}

\begin{figure}[htb]
\centering{%
\includegraphics[scale=0.4]{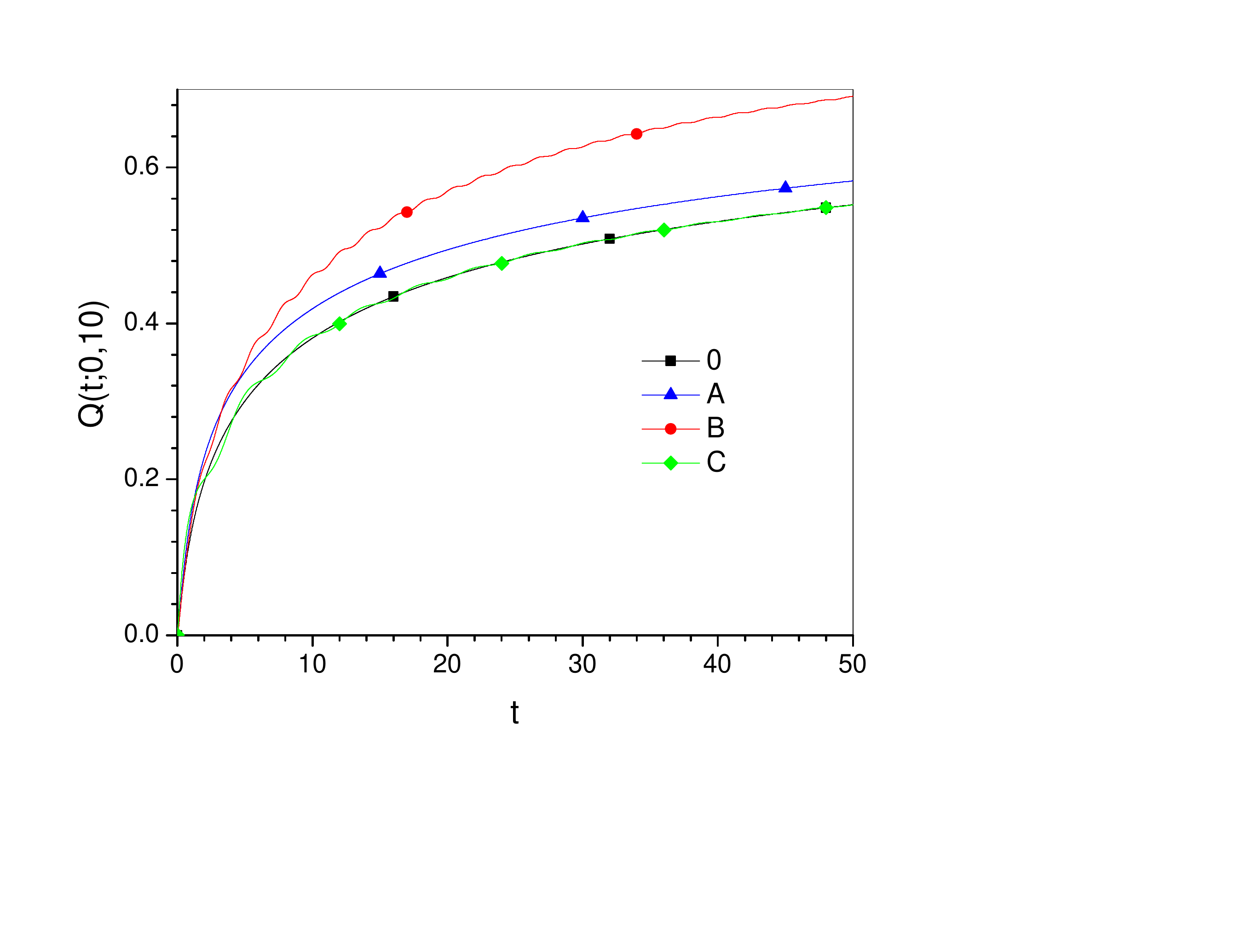}}
\caption{Time evolution of the function $Q$, the description is as in Fig. \ref{fig1}.}
\label{fig2}
\end{figure}

\begin{figure}[htb]
\centering{%
\includegraphics[scale=0.4]{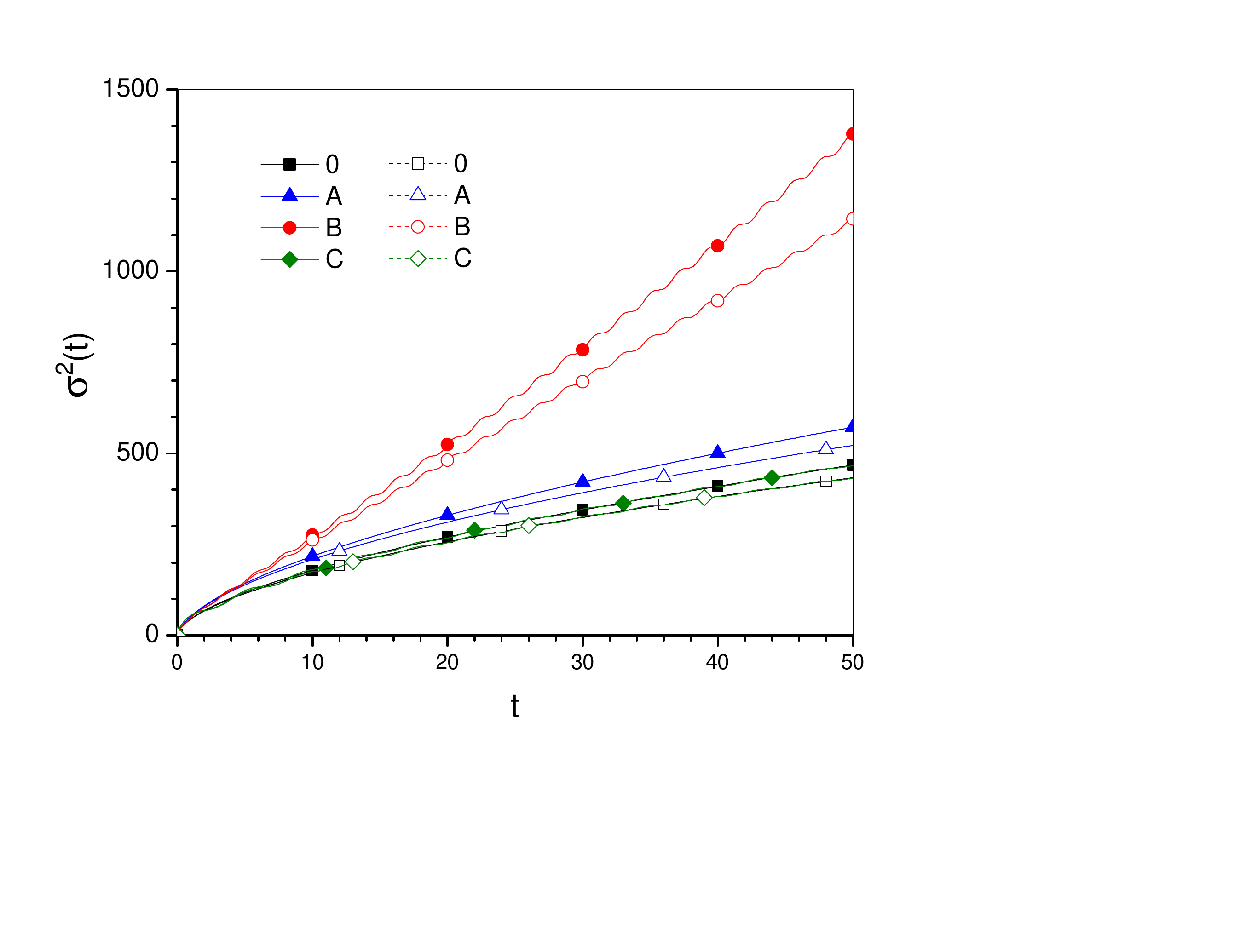}}
\caption{The plots of function $\sigma_\rho^2$ Eq. (\ref{eq15a}) for $\rho(t)\equiv 1$ (filled symbols) and for $\rho(t)$ varying in time (open symbols) are compared, the further description is as in Fig. \ref{fig1}.}
\label{fig3}
\end{figure}

\begin{figure}[htb]
\centering{%
\includegraphics[scale=0.4]{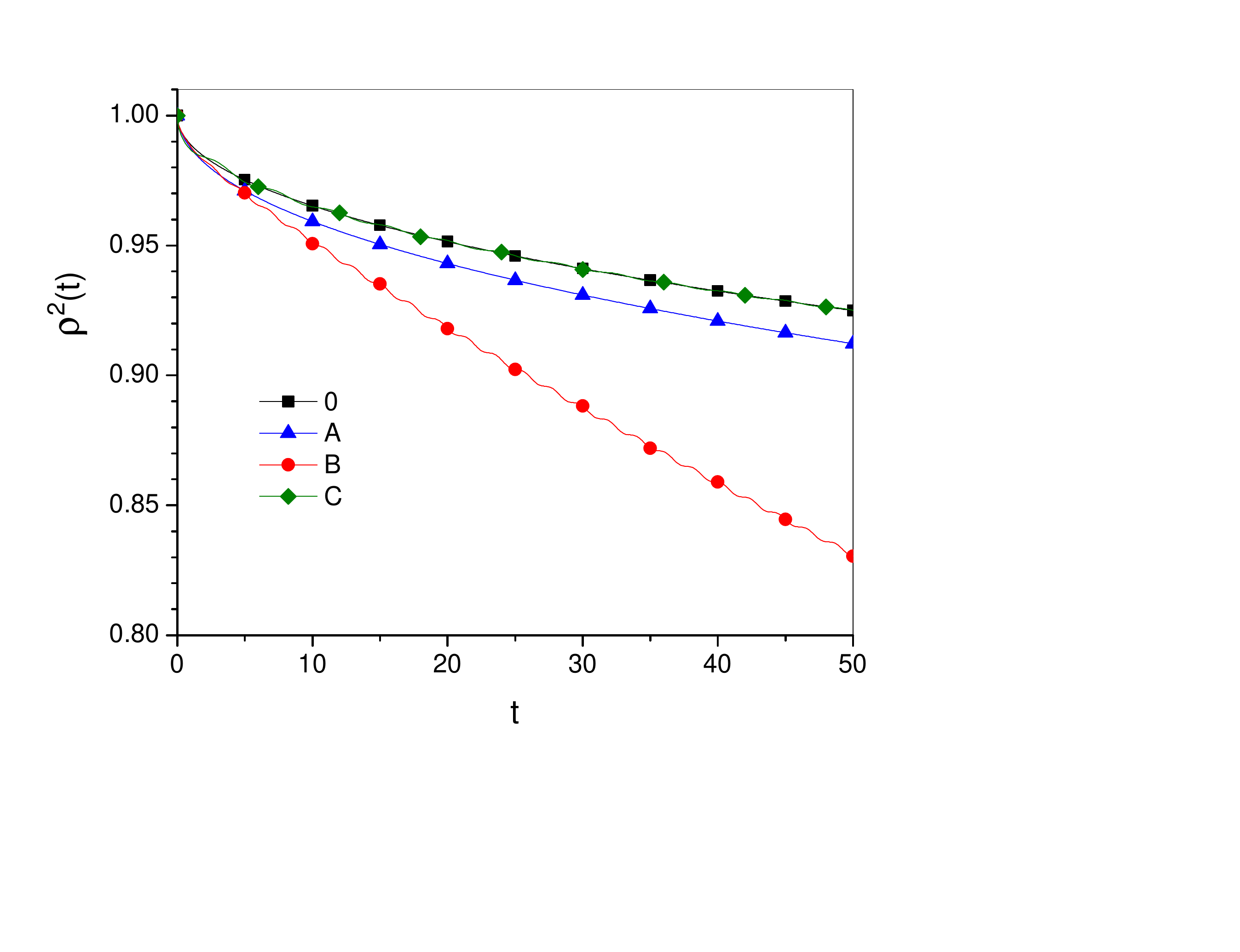}}
\caption{Time evolution of the probability that a particle still exists $\rho^2(t)$, the description is as in Fig. \ref{fig1}.}
\label{fig4}
\end{figure}

Fig. \ref{fig1} shows the plots of the first-passage time distribution Eq. (\ref{eq14}). In Fig. \ref{fig2} there is presented the time evolution of the probability that the molecule will reach the region $M_2$ by time $t$, which means that it has passed the point $x_M$ at least once. The function $Q$ has been numerically calculated from Eq. (\ref{eq15}). Examples of the functions $\sigma_\rho^2$ Eq. (\ref{eq15a}) and $\rho$ Eq. (\ref{eq10}) are shown in Figs. \ref{fig3} and \ref{fig4}.

The function $g_0$ does not change the timescale. In the cases $A$, $B$, and $C$, the timescale is changed and periodic functions are involved in the function $g$. In the case of $A$, $g_A$ contains ${\rm sin}(\tilde{\omega}(t))$, the period of this function decreases with time and goes to zero when $t\rightarrow\infty$. For the parameters given in the caption of Fig. \ref{fig1}, $g_A$ does not generate noticeable oscillating effects, see Fig. \ref{fig1}--\ref{fig4}. For processes $B$ and $C$ oscillation effects are observed in the figures. In process $B$, the amplitude of timescale oscillations increases with time and is proportional to $t$. The change in the timescale is relatively large, the rate of change is described by the function $g'_B(t)=1+at{\rm sin}^2(\omega t)$, which results in $g_B\sim t^{2\alpha}$ in the long time limit. Thus, assuming $\alpha>0.5$, the oscillation effect is superimposed here on the superdiffusion effect. In the case of $C$, the amplitude of oscillations are constant, the oscillation effect decreases with time. Here we have $\sigma_\rho^2(t)\sim t^\alpha$ when $t\rightarrow\infty$, there is the ``ordinary'' subdiffusion effect in the long time limit.

{\it Final remarks.} Subdiffusion and a molecule vanishing process are described by Eqs. (\ref{eq5}) and (\ref{eq6}). In general, involving $g$--Caputo fractional time derivatives in a model, instead of ``ordinary'' Caputo derivatives, we can describe changes of the course of the processes caused by external factors (e.g. temperature change causing the temporal evolution of parameters), and by the influence of different processes on each other. It has already been shown that the $g$--subdiffusion equation describe subdiffusion with a change of parameters $\alpha$ and $D$ \cite{kd2022} and a transition from "ordinary" subdiffusion to slow subdiffusion (ultraslow diffusion) \cite{kd2021a}. In this paper, we use the equations with $g$--Caputo fractional derivatives to model subdiffusion and molecule vanishing process which can be related to each other. 

We have assumed the molecule vanishing has a constant rate $\lambda$ and does not depend on the molecule position. This process can be interpreted as subdiffusion with $A\rightarrow\emptyset$ reaction. In \cite{kl} it was shown that the form of a subdiffusion--reaction equation depends on whether a diffusing molecule $A$ can decay with a probability independent of its position, or whether it must meet a $B$ molecule, with a certain probability, for a reaction to occur. Subdiffusion-reaction equations are qualitatively different in both cases \cite{kl,sokolov}. If the function $\rho$ also depended on the variable $x$ then the process could be described by a following $g$--subdiffusion--reaction equation with a reaction $nA+mB\rightarrow\emptyset$ ($n$ and $m$ are positive parameters), 
\begin{eqnarray}\label{eq16}
\frac{^C \partial^{\alpha}_{g_1} P_\rho(x,t|x_0)}{\partial t^\alpha}=D\frac{\partial^2 P_\rho(x,t|x_0)}{\partial x^2}\\
-\Xi(P_\rho, C_A(x,t),C_B(x,t)),\nonumber
\end{eqnarray}
where $P_\rho$ describes subdiffusion of $A$ molecule, $\Xi$ is a reaction term which can be controlled by both functions $g_1$ and $g_2$, $C_A$, $C_B$ denote concentrations of $A$ and $B$ molecules respectively. We note that if $n\neq 1$, a probability of the reaction depends on the concentration $C_A$. The reaction term may be non-linear, the analysis of a general $g$--subdiffusion--reaction equation will be presented elsewhere. $G$--subdiffusion with $A\rightarrow\emptyset$ reaction can be described by the subdiffusion equation and the molecule survival equation using the model presented in this paper.

\end{document}


\title{How to solve equations with the $g$--Caputo fractional derivative}

\author{Tadeusz Koszto{\l}owicz}
 \email{tadeusz.kosztolowicz@ujk.edu.pl}
 \affiliation{Institute of Physics, Jan Kochanowski University,\\
         Uniwersytecka 7, 25-406 Kielce, Poland}

\date{\today}

\maketitle

There are presented methods for solving subdiffusion equation and molecule survival equation, both with ``ordinary'' Caputo fractional time derivative and with $g$--Caputo fractional time derivative. All fractional derivatives are of the order belonging to the interval $(0,1)$. Solutions to equations with the ``ordinary'' Caputo derivative, and related functions, are marked with a tilde. Properties of $g$--Laplace transform and $g$--Caputo fractional derivatives can be found in \cite{almeida,jarad,fahad}.

\section{Solutions to subdiffusion equation}

The fractional subdiffusion equation with ``ordinary'' Caputo derivative of the order $\alpha\in(0,1)$ is
\begin{equation}\label{eq1}
\frac{^C \partial^{\alpha} \tilde{P}(x,t|x_0)}{\partial t^\alpha}=D\frac{\partial^2 \tilde{P}(x,t|x_0)}{\partial x^2},
\end{equation}
where the ``ordinary'' Caputo fractional derivative is defined for $0<\alpha<1$ as
\begin{equation}\label{eq2}
\frac{^Cd^{\alpha} \tilde{f}(t)}{dt^\alpha}=\frac{1}{\Gamma(1-\alpha)}\int_0^t (t-t')^{-\alpha}\tilde{f}'(t')dt',
\end{equation}
$\tilde{f}'(u)=d\tilde{f}(u)/du$, $\alpha$ is a subdiffusion parameter and $D$ is a generalized diffusion coefficient.

The Green's function $\tilde{P}(x,t|x_0)$ is the solution to subdiffusion equation for the initial condition 
\begin{equation}\label{eq3}
\tilde{P}(x,0|x_0)=\delta(x-x_0),
\end{equation}
where $\delta$ is the delta--Dirac function. For unbounded system the boundary conditions are
\begin{equation}\label{eq4}
\tilde{P}(-\infty,t|x_0)=\tilde{P}(\infty,t|x_0)=0.
\end{equation}

To solve Eq. (\ref{eq1}) the Laplace transform method can be used, the ``ordinary'' Laplace transform is defined as
\begin{equation}\label{eq5}
\mathcal{L}[\tilde{f}(t)](s)=\int_0^\infty {\rm e}^{-st}\tilde{f}(t)dt.
\end{equation}
Due to the relation
\begin{equation}\label{eq6}
\mathcal{L}\left[\frac{^C d^\alpha \tilde{f}(t)}{dt^\alpha}\right](s)=s^\alpha\mathcal{L}[\tilde{f}(t)](s)-s^{\alpha-1}\tilde{f}(0),
\end{equation}
$0<\alpha\leq 1$, we get from Eq. (\ref{eq1})
\begin{eqnarray}\label{eq7}
s^\alpha\mathcal{L}[\tilde{P}(x,t|x_0)](s)-s^{\alpha-1}\tilde{P}(x,0|x_0)\\
=D\frac{\partial^2\mathcal{L}[\tilde{P}(x,t|x_0)](s)}{\partial x^2}.\nonumber
\end{eqnarray}
In terms of the Laplace transform the boundary conditions read
\begin{equation}\label{eq8}
\mathcal{L}[\tilde{P}(-\infty,t|x_0)](s)=\mathcal{L}[\tilde{P}(\infty,t|x_0)](s)=0.
\end{equation}

Using the Fourier transform method, you can find the following solution to Eq. (\ref{eq5}) with the boundary conditions Eq. (\ref{eq8}) and the initial condition Eqs. (\ref{eq3})  
\begin{equation}\label{eq9}
\mathcal{L}[\tilde{P}(x,t|x_0)](s)=\frac{1}{2\sqrt{D}s^{1-\alpha/2}}\;{\rm e}^{-\frac{s^{\alpha/2}}{\sqrt{D}}|x-x_0|}.
\end{equation}
Using the formula \cite{tk2004}
\begin{eqnarray}\label{eq10}
\mathcal{L}^{-1}[s^\nu {\rm e}^{-as^\beta}](t)\equiv f_{\nu,\beta}(t:a)\\ 
=\frac{1}{t^{1+\nu}}\sum_{k=0}^\infty \frac{1}{k!\Gamma(-\nu-\beta k)}\left(-\frac{a}{t^\beta}\right)^k,\nonumber
\end{eqnarray}
where $a,\beta>0$, $\Gamma$ is the Gamma-Euler function, we obtain
\begin{equation}\label{eq11}
\tilde{P}(x,t|x_0)=\frac{1}{2\sqrt{D}}f_{-1+\alpha/2,\alpha/2}\left(t;\frac{|x-x_0|}{\sqrt{D}}\right).
\end{equation}
The function $f$ is the special case of the Wright function and the H-Fox function.

Let the function $g$, given in a time unit, fulfils the conditions $g(0)=0$, $g(\infty)=\infty$, and $g(t),g'(t)>0$ for $t>0$.
The Caputo fractional derivative of the order $\alpha\in(0,1)$ with respect to the function $g$ is defined as 
\begin{equation}\label{eq12}
\frac{^Cd^{\alpha}_g f(t)}{dt^\alpha}=\frac{1}{\Gamma(1-\alpha)}\int_0^t (g(t)-g(u))^{-\alpha}f'(u)du.
\end{equation}
When $g(t)=t$, the $g$-Caputo fractional derivative takes the form of the ``ordinary'' Caputo derivative (\ref{eq2}).

The $g$-subdiffusion equation is defined as 
\begin{equation}\label{eq13}
\frac{^C \partial^{\alpha}_g P(x,t|x_0)}{\partial t^\alpha}=D\frac{\partial^2 P(x,t|x_0)}{\partial x^2}.
\end{equation}
To solve Eq. (\ref{eq13}) it is convenient to use the Laplace transform with respect to another function $g$ (the $g$--Laplace transform)
\begin{equation}\label{eq14}
\mathcal{L}_g[f(t)](s)=\int_0^\infty {\rm e}^{-s g(t)}f(t)g'(t)dt.
\end{equation}
The Laplace transforms are related to each other by the relation
\begin{equation}\label{eq15}
\mathcal{L}_g[f(t)](s)=\mathcal{L}[\tilde{f}(g^{-1}(t))](s).
\end{equation}
Eq. (\ref{eq15}) and the Lerch's uniqueness of the inverse Laplace transform theorem provide the following rule
\begin{equation}\label{eq16}
\mathcal{L}_g[f(t)](s)=\mathcal{L}[\tilde{f}(t)](s)\Leftrightarrow f(t)=\tilde{f}(g(t)).
\end{equation}
The above relation is the basis of the method of solving the $g$--subdiffusion equation and is helpful in calculating the inverse $g$--Laplace transform if the inverse "ordinary" Laplace transform is known. Using the formula
\begin{equation}\label{eq17}
\mathcal{L}\left[t^\nu\right](s)=\frac{\Gamma(1+\nu)}{s^{1+\nu}},\;\nu>-1,
\end{equation}
and Eq. (\ref{eq16}) we get
\begin{equation}\label{eq18}
\mathcal{L}_g^{-1}\left[\frac{1}{s^{1+\nu}}\right](t)=\frac{g^\nu(t)}{\Gamma(1+\nu)}\;,\;\nu>-1.
\end{equation}
From Eqs. (\ref{eq10}) and (\ref{eq16}) we obtain
\begin{eqnarray}\label{eq19}
\mathcal{L}_g^{-1}[s^\nu {\rm e}^{-as^\mu}](t)\equiv f_{\nu,\mu}(g(t);a)\\
=\frac{1}{g^{1+\nu}(t)}\sum_{k=0}^\infty \frac{1}{k!\Gamma(-\nu-\mu k)}\left(-\frac{a}{g^\mu(t)}\right)^k,\nonumber
\end{eqnarray}
$a,\mu>0$.

The $g$--Laplace transform has the following property that makes the procedure for solving Eq. (\ref{eq13}) similar to the procedure for solving Eq. (\ref{eq1}) using the "ordinary" Laplace transform
\begin{equation}\label{eq20}
\mathcal{L}_g\left[\frac{^Cd^{\alpha}_g f(t)}{dt^\alpha}\right](s)=s^\alpha\mathcal{L}_g[f(t)](s)-s^{\alpha-1}f(0).
\end{equation}
Due to Eq. (\ref{eq20}), in terms of the $g$--Laplace transform the $g$--subdiffusion equation is
\begin{eqnarray}\label{eq21}
s^\alpha\mathcal{L}_g[P(x,t|x_0)](s)-s^{\alpha-1}P(x,0|x_0)\\
=D\frac{\partial^2\mathcal{L}_g[P(x,t|x_0)](s)}{\partial x^2}.\nonumber
\end{eqnarray}
The structure of Eq. (\ref{eq21}) as a differential equation with respect to $x$ variable is the same as the structure of Eq. (\ref{eq7}). The solution to Eq. (\ref{eq21}) for the boundary conditions 
\begin{equation}\label{eq22}
\mathcal{L}_g[P(-\infty,t|x_0)](s)=\mathcal{L}_g[P(\infty,t|x_0)](s)=0
\end{equation}
and the initial condition
\begin{equation}\label{eq23}
P(x,0|x_0)=\delta(x-x_0)
\end{equation}
is
\begin{equation}\label{eq24}
\mathcal{L}_g[P(x,t|x_0)](s)=\frac{1}{2\sqrt{D}s^{1-\alpha/2}}\;{\rm e}^{-\frac{s^{\alpha/2}}{\sqrt{D}}|x-x_0|}.
\end{equation}
From Eqs. (\ref{eq9}) and (\ref{eq24}) we obtain
\begin{equation}\label{eq25}
\mathcal{L}_g[P(x,t|x_0)](s)=\mathcal{L}[\tilde{P}(x,t|x_0)](s).
\end{equation}
Due to the relation Eq. (\ref{eq16}) we have
\begin{equation}\label{eq26}
P(x,t|x_0)=\tilde{P}(x,g(t)|x_0).
\end{equation}
Finally, we get
\begin{equation}\label{eq27}
P(x,t|x_0)=\frac{1}{2\sqrt{D}}f_{-1+\alpha/2,\alpha/2}\left(g(t);\frac{|x-x_0|}{\sqrt{D}}\right).
\end{equation}
We note that the same results can be obtained from Eq. (\ref{eq19}) and (\ref{eq24}).

\section{Solutions to the molecule survival equation}

The ``ordinary'' fractional molecule survival equation is defined as
\begin{equation}\label{eq28}
\frac{^C d^\beta \tilde{\rho}(t)}{dt^\beta}=-\lambda\tilde{\rho}(t),
\end{equation}
where a positive parameter $\lambda$ is a molecule vanishing rate. The initial condition is
\begin{equation}\label{eq29}
\tilde{\rho}(0)=1.
\end{equation}
In order to solve Eq. (\ref{eq28}) we use the relations
\begin{equation}\label{eq30}
\frac{^C d^\beta t^r}{dt^\beta}=\frac{\Gamma(r+1)}{\Gamma(r-\beta+1)}t^{r-\beta},\; r>[\beta],
\end{equation}
\begin{equation}\label{eq31}
\frac{^C d^\beta 1}{dt^\beta}=0,
\end{equation}
and the Mittag--Leffler function
\begin{equation}\label{eq32}
{\rm E}_\beta(z)=\sum_{n=0}^\infty \frac{z^n}{\Gamma(\beta n+1)}.
\end{equation}
$\beta>0$.
From Eqs. (\ref{eq30})--(\ref{eq32}), differentiating term by term the series in Eq. (\ref{eq32}), we obtain
\begin{equation}\label{eq33}
\frac{^C d^\beta {\rm E}_\beta(-\lambda t^\beta)}{dt^\beta}=-\lambda {\rm E}_\beta(-\lambda t^\beta).
\end{equation}
Comparing Eqs. (\ref{eq28}) and (\ref{eq33}) we get
\begin{equation}\label{eq34}
\tilde{\rho}(t)={\rm E}_\beta(-\lambda t^\beta).
\end{equation}
From Eqs. (\ref{eq6}), (\ref{eq28}), and (\ref{eq29}), we find 
\begin{equation}\label{eq35}
\mathcal{L}[\tilde{\rho}(t)](s)=\frac{s^{\beta-1}}{s^\beta +\lambda}.
\end{equation}
We note that from Eqs. (\ref{eq34}) and (\ref{eq35}) we obtain
\begin{equation}\label{eq36}
\mathcal{L}[{\rm E}_\beta(-\lambda t^\beta)](s)=\frac{s^{\beta-1}}{s^\beta +\lambda}.
\end{equation}
The approximation of the small parameter $s$ corresponds to the long time approximation of the inverse Laplace transform. From Eqs. (\ref{eq17}) and (\ref{eq35}) we get
\begin{equation}\label{eq37}
\tilde{\rho}(t)\approx\frac{1}{\lambda\Gamma(1-\beta) t^\beta}
\end{equation}
when $t\rightarrow\infty$.

We  assume the following form of the $g$--molecule survival equation
\begin{equation}\label{eq38}
\frac{^C d^\beta_{g}\rho(t)}{dt^\beta}=-\lambda\rho(t).
\end{equation}
In terms of $g$--Laplace transform, the solution to Eq. (\ref{eq38}) for the initial condition $\rho(0)=1$ is 
\begin{equation}\label{eq39}
\mathcal{L}_g[\rho(t)](s)=\frac{\lambda s^{\beta-1}}{s^\beta+\lambda}.
\end{equation}
From Eqs. (\ref{eq16}), (\ref{eq34}), (\ref{eq35}), and (\ref{eq39}) we get
\begin{equation}\label{eq40}
\rho(t)={\rm E}_\beta(-\lambda g^\beta(t)).
\end{equation}